\documentclass[graybox]{svmult}

\usepackage{helvet}         % selects Helvetica as sans-serif font
\usepackage{courier}        % selects Courier as typewriter font
\usepackage{type1cm}        % activate if the above 3 fonts are
\usepackage{makeidx}         % allows index generation
\usepackage{graphicx}        % standard LaTeX graphics tool
\usepackage{multicol}        % used for the two-column index
\usepackage[bottom]{footmisc}% places footnotes at page bottom
\usepackage{amsmath,amssymb}
\usepackage{mathrsfs}
\begin{document}

\title*{Bulk reconstruction from a scalar CFT at the boundary by the smearing with the flow equation
}
\titlerunning{Bulk reconstruction by the smearing with the flow equation}
\author{Sinya Aoki, Janos Balog, Tetsuya Onogi and Shuichi Yokoyama}
\authorrunning{S. Aoki, J. Balog, T. Onogi and S. Yokoyama} 
\institute{Sinya Aoki \at Yukawa Institute for Theoretical Physics, Kyoto University, Sakyo-ku, Kyoto 606-8502, Japan, \email{saoki@yukawa.kyoto-u.ac.jp}
\and Janos Balog \at Institute for Particle and Nuclear Physics, Wigner Research Center for Physics, H-1525 Budapest 114, P.O.B. 49, Hungary, \email{balog.janos@wigner.hu}
\and Tetsuya Onogi \at Department of Physics, Osaka University, Toyonaka, Osaka 560-0045, Japan, \email{onogi@het.phys.sci.osaka-u.ac.jp}
\and Shuichi Yokoyama \at Yukawa Institute for Theoretical Physics, Kyoto University, Sakyo-ku, Kyoto 606-8502, Japan, \email{shuichi.yokoyama@yukawa.kyoto-u.ac.jp}
}
\maketitle

\abstract{We explain our proposal for an alternative bulk reconstruction of AdS/CFT correspondences from a scalar field by the flow method.
By smearing and then normalizing a primary field in a $d$ dimensional CFT, we construct  a  bulk field, through which a $d+1$ dimensional AdS space emerges. }

\section{Introduction}
\label{sec:1}

The AdS/CFT correspondence\cite{Maldacena:1997re} is a key to understand a holographic nature of gravity and 
may give a hint for quantum gravity.
Although a plenty of circumstance evidences exist, an essential mechanism why AdS/CFT correspondence realizes
has not been fully established yet.
While the correspondence may be explained by the string theory, an alternative but more universal mechanism might exist.

One of key questions is how an additional dimension of the AdS emerges from the CFT at the boundary.
One approach, called the HKLL bulk reconstruction\cite{Hamilton:2005ju,Hamilton:2006az},  
gives a relation between a bulk local operator in the AdS and CFT operators at the boundary with Lorentizan signature.

Recently we have proposed an alternative bulk reconstruction by the so-called flow method by the path-integral with Euclidean signature\cite{Aoki:2017bru,Aoki:2017uce,Aoki:2018dmc,Aoki:2019bfb,Aoki:2019dim,Aoki:2020ztd},
which is explained in this talk.

\section{Bulk reconstruction by the flow}
\label{sec:2}
\subsection{From conformal symmetry to bulk symmetry}
Let us consider a non-singlet primary field of a conformal field theory (CFT) on a $d$ dimensional Euclidean space, whose 2-pt function behaves as
\begin{equation}
\langle \varphi^a(x) \varphi^b(y) \rangle = \delta^{ab} {1\over \vert x-y\vert^{2\Delta}}, 
\label{eq:2pt_CFT}
\end{equation}
where $\Delta$ is a conformal dimension of $\varphi^a$ and $a,b\cdots$ represent indices of a global symmetry such as an $O(N)$ symmetry.

We first make one parameter extension of the field $\varphi^a$ via the generalized flow equation as
\begin{eqnarray}
(-\alpha \eta \partial_\eta^2 +\beta\partial_\eta) \phi^a(x,\eta) = \Box \phi^a(x,\eta), \quad
\phi^a(x,0)=\varphi^a(x),
\label{eq:flow}
\end{eqnarray}
where $\alpha,\beta$ are parameters to control this extension. We regard this process as the smearing  of the field $\varphi^a$, since the flow equation becomes the heat equation at $\alpha=0$. 
We then normalize $\phi^a$ as
\begin{eqnarray}
\sigma^a(X) =  {\phi^a(x,\eta) \over \sqrt{ \langle \phi(x,\eta)\cdot\phi(x,\eta) \rangle}} \Rightarrow \langle \sigma (X)\cdot\sigma (X) \rangle =1,  
\end{eqnarray}
where the average is taken for the CFT in $d$ dimensions as in \eqref{eq:2pt_CFT},  $X:=(z,x)$ is a $d+1$ dimensional coordinate with $\eta:= \alpha z^2/4$, and
$F\cdot G :=\sum_a F^a G^a$. The smearing followed by normalization may be interpreted as a kind of the renormalization group transformation.
The field $\sigma^a$ is expressed in the smeared form by the integral as 
\begin{eqnarray}
\sigma^a(X) = \int d^dy\, h(z,x-y) \varphi^a(y).
\end{eqnarray}

The smearing kernel $h(z,x)$ is determined by conditions that the conformal transformation $U$ to $\varphi^a$, given by
\begin{eqnarray}
U \varphi^a(y) U^\dagger := \tilde\varphi^a(y) =J^\Delta(y) \varphi^a(\tilde y), 
\end{eqnarray}
generates the coordinate transformation to  $\sigma^a$ in $d+1$ dimensions as
\begin{eqnarray}
U \sigma^a(X) U^\dagger := \tilde\sigma^a(X) =\sigma^a(\tilde X), 
\end{eqnarray}
where
\begin{description}
\item[1. translation] $\tilde y^\mu = y^\mu+a^\mu$, $J(y)=1$, $\tilde x^\mu=x^\mu+a$, $\tilde z=z$,
\item[2. rotation] $\tilde y^\mu = \Omega^\mu{}_\nu y^\nu$, $J(y)=1$, $\tilde x^\mu=  \Omega^\mu{}_\nu x^\nu+a$, $\tilde z=z$,
\item[3. dilatation]  $\tilde y^\mu = \lambda y^\mu$, $J(y)=\lambda$, $\tilde X^A=\lambda X^A $,
\item[4. inversion]  $\tilde y^\mu = y^\mu/y^2$, $J(y)=1/y^2$, $\tilde X^A= X^A/X^2 $,
\end{description}
which generate SO$(d+1,1)$ transformation. 
 Transformations 1--3 imply
\begin{equation}
\sigma^a(X) = z^{\Delta-d}\int d^dy\, \Sigma\left( 1+{(x-y)^2\over z^2} \right) \varphi^a(y),
\end{equation}
where $\Sigma$ is an arbitrary function.
The transformation 4 to the CFT operator in the above equation generates
\begin{eqnarray}
&&z^{\Delta -d} \int d^dy\, \Sigma\left( 1+{(x-y)^2\over z^2} \right) {\varphi(q=\tilde y)\over (y^2)^\Delta}
=  \int d^dq\, \Sigma\left( 1+{(x-\tilde q)^2\over z^2} \right) (zq^2)^{\Delta-d} \varphi(q)\nonumber \\
&=& \left({\tilde z\over \tilde X^2}\right)^{\Delta-d} \int d^dq\, \Sigma\left( {\tilde X^2\over q^2}\left[1+{(\tilde x-q)^2\over \tilde z^2}\right] \right) (q^2)^{\Delta-d} \varphi(q),
\end{eqnarray}
while the bulk operator transforms as
\begin{eqnarray}
\sigma^a(\tilde X) &=& \tilde z^{\Delta-d} \int  d^dq\, \Sigma\left( 1+{(\tilde x-q)^2\over \tilde z^2} \right)  \varphi(q).
\end{eqnarray}
Thus the symmetry implies $\Sigma(u) = \Sigma_0 u^{\Delta-d}$, which finally gives
\begin{equation}
\sigma^a(X) = \int d^dy\, h(z,x-y) \varphi^a(y),
\quad h(z,x) = \Sigma_0 \left( {z\over z^2+x^2} \right)^{d-\Delta},
\end{equation}
where $\Delta < d$ is necessary for this expression to be regarded as a smearing.
After the Wick rotation,
this smearing kernel is identical to the one in the HKLL\cite{Hamilton:2006az}, though
$\Delta > d-1$ is required for the HKLL.

\subsection{Some properties}
It is easy to see the kernel $h$ satisfies $(\Box_{\rm AdS} -m^2) h(X) =0$, where $\Box_{\rm AdS} := z^2(\partial_z^2+\Box) -(d-1)z \partial_z$ and 
$m^2 R^2 := (\Delta -d) \Delta$ with some length scale $R$, which turns out to be the AdS radius. 
Furthermore, $h(X)$ corresponds to a solution to the flow equation \eqref{eq:flow}  with $\beta/\alpha =(d-2)/2-\Delta$.
As a result, $\sigma^a(X)$ satisfies equations of motion for a scalar field in the Euclidean AdS space as
\begin{equation}
(\Box_{\rm AdS} - m^2) \sigma^a(X) = 0.
\end{equation}
 
Since,  for $\Delta < d/2$, 
\begin{equation}
\lim_{z\to 0} {z^{d-2\Delta}\over (x^2+z^2)^{d-\Delta}} \sim \left\{
\begin{array}{cc}
  z^{d-2\Delta} \to 0, & x\not=0 ,   \\
  z^{-d} \to \infty, & x=0,    \\
\end{array}
\right.
\end{equation}
\begin{equation}
\int d^dx\, {z^{d-2\Delta}\over (x^2+z^2)^{d-\Delta}} ={\pi^{d/2}\Gamma(d/2-\Delta)\over \Gamma(d-\Delta)}:={1\over \Lambda},
\end{equation}
we can write
\begin{equation}
\lim_{z\to 0} {z^{d-2\Delta}\over (x^2+z^2)^{d-\Delta}} = {\delta^{(d)}(x)\over \Lambda} .
\end{equation}
Thus the smearing function satisfies
\begin{equation}
\lim_{z\to 0} h(z,x) =  {\Sigma_0\over \Lambda} z^{\Delta}\delta^{(d)}(x),
\end{equation}
which leads to a BDHM relation\cite{Banks:1998dd} as
\begin{equation}
\lim_{z\to 0} z^{-\Delta} \sigma^a(X) = {\Sigma_0\over \Lambda} \varphi^a(x).
\end{equation}

Using a singlet bulk composite scalar field $S(x) :=\sum_a \sigma^a(X) \sigma^a(X)$, 
we can define a bulk to boundary correlation function as
\begin{equation}
{\cal F}_O (X,y) := \langle S(X) O(y)\rangle
\end{equation}
where $O(y)$ is an arbitrary singlet scalar primary field  with a conformal dimension $\Delta_O$ at the boundary.
A combination of the conformal symmetry and the bulk symmetry in the previous subsection leads to
 \begin{equation}
{\cal F}_O (X,y) = C_O \left( {z\over (x-y)^2+z^2}\right)^{\Delta_O},
\end{equation}
which is exact in the sense that only unknown constant $C_O$ depends on the detail of the boundary CFT such as coupling constants,
and reproduces the standard prediction by the AdS/CFT correspondence. 
Furthermore, ${\cal F}$ satisfies
\begin{equation}
(\Box_{\rm ADS}^X -m_O^2) {\cal F}(X,y) = 0, \quad m_O^2 := {\Delta_O (\Delta_O -d)\over R^2 }.
\end{equation}

\section{Symmetries and bulk geometry}
In this section we investigate what kind of space the $d+1$ dimensional coordinate $X$ describes.

\subsection{Constraints to a generic correlation function}
Let us consider a generic correlation function, which contains $m$ bulk fields and $s$ boundary fields, given by
\begin{equation}
\left\langle \prod_{i=1}^m G^i_{A_1^iA_2^i\cdots A^i_{n_i}} (X_i)\prod_{j=1}^s O^j_{\mu_1^j\mu_2^j\cdots \mu_{l_j}^j} (y_j)\right\rangle ,
\end{equation}
where $A_1^i A_2^i \cdots A^i_{n_i}$ is a tensor index of a bulk operator $G^i$, while
$\mu_1^j\mu_2^j\cdots \mu_{l_j}^j$ is that for a boundary operator $O^j$. 
The conformal symmetry at the boundary with the associated coordinate transformation in the bulk gives an exact quantum relation as
\begin{eqnarray}
\left\langle \prod_{i=1}^m G^i_{A_1^iA_2^i\cdots A^i_{n_i}} (\tilde X_i)\prod_{j=1}^s O^j_{\mu_1^j\mu_2^j\cdots \mu_{l_j}^j} (\tilde y_j)\right\rangle 
= \prod_{i=1}^m {\partial X_i^{B_1^i}\over \partial \tilde X_i^{A_1^i}} {\partial X_i^{B_2^i}\over \partial \tilde X_i^{A_2^i}}
\cdots {\partial X_i^{B_{n_i}^i}\over \partial \tilde X_i^{A_{n_i}^i}}\nonumber \\
\times\prod_{j=1}^s J(y_j)^{-\Delta_j}{\partial y_j^{\nu_1^j}\over \partial \tilde y_j^{\mu_1^j}} 
 {\partial y_j^{\nu_2^j}\over \partial \tilde y_j^{\mu_2^j}} \cdots {\partial y_j^{\nu_{l_j}^j}\over \partial \tilde y_j^{\mu_{l_j}^j}} 
\left\langle \prod_{i=1}^m G^i_{B_1^iB_2^i\cdots B^i_{n_i}} (X_i)\prod_{j=1}^s O^j_{\nu_1^j\nu_2^j\cdots \nu_{l_j}^j} ( y_j)\right\rangle . \nonumber \\
\label{eq:const_general}
\end{eqnarray}

\subsection{Metric field}
We define the bulk metric field as\cite{Aoki:2015dla}
\begin{equation}
g_{AB}(X) := \ell^2\sum_{a=1}^N \partial_A\sigma^a(X) \partial_B\sigma^a(X), 
\end{equation}
which is the simplest among singlet and symmetric 2nd rank tensors in the bulk,   
where $\ell$ is some length scale. 
It is worth noting that this metric field is finite without ultra-violate (UV) divergence thank to the smearing.

The metric becomes classical in the large $N$ limit due to the large $N$ factorization as\cite{Aoki:2015dla}
\begin{equation}
\langle g_{AB}(X_1) g_{CD} (X_2)\rangle = \langle g_{AB}(X_1)\rangle \langle g_{CD} (X_2)\rangle + O(1/N),  
\end{equation}
which make a vacuum expectation value (VEV) of the quantum Einstein tensor $G_{AB}$ classical:
\begin{eqnarray}
\langle G_{AB} ( g_{CD} ) \rangle = G_{AB} ( \langle g_{CD}\rangle ) + O(1/N). 
\end{eqnarray}
A classical geometry appears after quantum averages.

The VEV of the metric can be interpreted as the (Bures) information metric\cite{Aoki:2017bru}, which defines a distance between (mixed state) density matrices $\rho$ and $\rho +d\rho$ as
\begin{equation}
d^2(\rho,\rho+d\rho) := {1\over 2} {\rm tr} [ d\rho\, G], 
\end{equation}
where $G$ satisfies $ \rho G + G\rho = d\rho$. In our case, $\rho$ is given by a mixed state as
\begin{equation}
\rho(X) := \sum_{a=1}^N \vert \sigma^a(X) \rangle \langle \sigma^a(X)\vert,
\end{equation}
which represents $N$ entangled pairs, where the inner product between states is defined by $\langle \sigma^a(X) \vert \sigma^b(Y)\rangle =
\delta^{ab} \langle \sigma(X)\cdot \sigma(Y)\rangle/N$. 
Using this definition, we obtain
\begin{equation}
\ell^2 d^2(\rho,\rho+d\rho) = \langle g_{AB}(X) \rangle dX^A dX^B.
\end{equation}
Thus the VEV of the metric field $g_{AB}(X)$ defines a distance in the bulk space through $d^2(\rho,\rho+d\rho)$ in unit of $\ell^2$.

\subsection{VEV of the metric field} 
Let us calculate the VEV of the metric field. The constraint \eqref{eq:const_general} applied to $g_{AB}$ leads to
\begin{equation}
\langle 0\vert g_{AB}(X) \vert 0\rangle = R^2{\delta_{AB}\over z^2}, 
\end{equation}
which describes the AdS space in the Poincare coordinate, where an unknown constant $R$ is the radius of the AdS space.
 The explicit form of the 2-pt function of  CFT in \eqref{eq:2pt_CFT} gives $R^2 =\ell^2\Delta(d-\Delta)/(d+1)$, which is positive since $\Delta< d/2$.
The boundary CFT generates the bulk AdS, and the bulk symmetry in the previous section is equal to the AdS isometry on
$\langle 0\vert g_{AB}(X) \vert 0\rangle$.  Thus the AdS/CFT correspondence is realized by the flow construction.

\subsection{Scalar excited state contribution}  
We finally consider how excited states modify the AdS structure described by the VEV of the metric field.
As the simplest example, we  calculate the VEV of the metric in the presence of the source $J$ coupled to the singlet CFT scalar field at the origin
in the radial quantization as
\begin{equation}
\bar g_{AB}(X)=\langle 0 \vert g_{AB}(X)e^{J O(0)}\vert 0\rangle =
\langle 0 \vert g_{AB}(X)\vert 0\rangle + J\langle 0 \vert g_{AB}(X)\vert S\rangle +O(J^2),
\end{equation}
where we need to calculate
\begin{equation}
\langle 0 \vert g_{AB}(X)\vert S\rangle = \lim_{y^2\to 0} G_{AB}(X,y), \quad
G_{AB}(X,y) := \langle 0 \vert g_{AB}(X) O(y)\vert 0\rangle.
\end{equation}
The constraint \eqref{eq:const_general} to $G_{AB}$ reads
\begin{eqnarray}
J(y)^{\Delta_O} G_{AB}(X,y) ={\partial \tilde X^C\over \partial X^A} {\partial \tilde X^D\over \partial X^B} G_{CD}(\tilde X, \tilde y), 
\end{eqnarray}
which leads to
\begin{eqnarray}
G_{AB}(X,y) =T^{\Delta_O}(X,y) \left[a_1{\delta_{AB}\over z^2} + a_2 {T_A(X,y) T_B(X,y)\over T^2(X,y)}\right], 
\end{eqnarray}
where $a_1$ and $a_2$ are unknown constants, and 
\begin{equation}
T(X,y) := {z\over (x-y)^2+z^2}, \quad T_A(X,y):= \partial_A T(X,y).
\end{equation}
We thus obtain
\begin{eqnarray}
\bar g_{AB}(X) ={\delta_{AB}\over z^2} \left[R^2 + a_1 J \left({z\over x^2+z^2}\right)^{\Delta_O}\right]
+ a_2 J T_A T_B \left({z\over x^2+z^2}\right)^{\Delta_O-2} ~~~~~
\end{eqnarray}
at the 1st order in $J$,
where 
\begin{equation}
T_z:= {x^2-z^2\over (x^2+z^2)^2},\quad
T_\mu := - { 2x_\mu z\over  (x^2+z^2)^2}.
\end{equation}
This metric describes the asymptotically AdS space since
\begin{equation}
 \bar g_{AB}(X)  = R^2 {\delta_{AB}\over z^2}\left[1 + O\left( z^{\Delta_O}\right)\right],
\end{equation}
as $z\to 0$.

\section{Summary and discussion}
Smearing and normalizing the non-singlet primary field $\varphi^a$ with $\Delta <d/2$ at the boundary give the bulk field $\sigma^a$.
Through this process, the conformal symmetry turns into the bulk coordinate transformation, which leads to the expected behavior for the bulk to boundary correlation function.   As the VEV of the metric field describes the AdS space, the AdS/CFT correspondence is naturally realized by our method.

As one of future problems, let us consider a bulk to bulk correlation function for the scalar, which is evaluated as
\begin{eqnarray}
\langle S(X_1) S(X_2) \rangle &=& 1 +{2\over N} G^2(X_1,X_2) + \langle S(X_1) S(X_2) \rangle_c, 
\end{eqnarray}
where
\begin{equation}
G(X_1,X_2) = {}_2 F_1\left({\Delta\over 2}, {d -\Delta\over 2};  {d+1 \over 2}; 1-{R_{12}^2\over 4}\right), 
R_{12} :={(x_1-x_2)^2+z_1^2+z_2^2\over z_1 z_2},
\end{equation}
which is non-singular at $X_1=X_2$.  Since the connected part given by the last term is absent for the free CFT,
the corresponding bulk theory is non-local (stringy).
Thus the interaction in the CFT is required to recover the locality of the bulk theory.

%\bigskip
%
\begin{acknowledgement}
This work is supported in part by the Grant-in-Aid of the Japanese Ministry of Education, Sciences and Technology, Sports and Culture (MEXT) for Scientific Research (Nos.~JP16H03978,  JP18K03620, JP18H05236, JP19K03847), and by the NKFIH grant K134946.
\end{acknowledgement}
\end{document}